\documentclass[12pt]{amsart}
\usepackage{graphicx}
\usepackage{epsfig}
\usepackage{fancyhdr}
\usepackage{framed}
\usepackage{floatrow}
\usepackage{mdframed}

\usepackage{amsmath,amsfonts,amssymb,amsthm,mathabx}
\theoremstyle{plain}

\newtheorem*{thm*}{Theorem}

\newtheorem*{lem*}{Lemma}

\newtheorem*{cor*}{Corollary}

\newtheorem*{prop*}{Proposition}

\theoremstyle{definition}

\def\\{{\mathcal{R}}}

\def\be{\begin{equation}}
\def\ee{\end{equation}}
\def\ben{\begin{equation*}}
\def\een{\end{equation*}}

\begin{document}

\title{Modeling of human society as a locally interacting product-potential networks of automaton}
\author{Vladislav B. Kovchegov}
\email{vlad\_kovchegov@yahoo.com}
\begin{abstract}  
 The central problems in social sciences concern the social and psychological
  mechanisms and conditions required for the emergence and stability of human groups.
  The present article is dedicated to the problem of stability of human groups.
  We model human groups using local interacting systems of automaton with relations
  and reactions and using the structural balance theory. The `structural balance
  theory' ties the emergence of a human group with the human actor's thoughts about
  how another actor treats him and his perception of actors. The Cartwright and
  Harary formalization the concept of balance theory within a graph theoretical
  setting unable to get a number of mathematical results pertaining to an algebraic
  formulation of the theory of balance in signed networks/graphs. The deeper
  generalization of 'balance theory' as the smooth product-potential fields on
  domain gives us the ability to create theory of 'smooth product potential social
  fields'. We then find that all discrete product-potential system tightly connect
  with other process - process multiplication on the randomly chosen matrices and
  we find connections between stationary measures and some algebraic objects.
\end{abstract}

\date{}
\maketitle


 \centerline{ \bf Introduction }
\par The present article is dedicated to the problem of stability of human groups,
firstly by modeling human group using locally interacting systems
of automaton with relations and reactions. We will concentrate our
attention on the particular type of local interacting systems of
automata named the product potential social
 system. The product potential social system is a mathematical analogy of
 balance in the social sense human group. The problem of stability has a social source.
 The source of mathematical realization is the cellular automata theory and the theory
 of interacting systems.
\par {\bf Homans and group dynamics(1958,1974).}  Homans defines the social structure
of the group as an equilibrium state of social system (he defines
the term "group", too). This is a good way of linking dynamics and
structure. He wants to specify the mechanism that produce and
maintain a social structure. He speculates that the process would
be described by the system of differential equations that would
have equilibrium solutions and these solution would be those
states of social system that are the social structure.
\par  He distinguishes an "internal system" and "external system". It means that
any social system has two functional problem: adapting to an
environment and integrating its units. He asserted that the more
frequently persons interact, the more similar they become in
sentiments. But he had not really modelled the process of social
interaction that would yield this property. To rectified this
 deficiency Homans created the conception of  "social behavior as exchange".
Social interaction is an exchange involving  such the key
variables as value of activities and frequency of activity. The
Homan's exchange theory and an exchange in the competitive market
economics are based on the very same behavioral exchange.
\par  Coleman had studied the exchange formulated by Homans and he (1990) produced
his own version of this idea. This work uses a general equilibrium
framework from competitive market economics as it did Homans.
Coleman treats problem such as emergence of norms, interpersonal
comparison of utilities, and the derivation of the interests of
collectivity from the exchange among its part. Coleman's model
start from two relations connecting two sets of entities: actors
and resources (events). Each actor has some level of interest in
each resource and also some proportion of control over each
resource. Actors are willing to give up some control over some
resources to gain more control over other resources of interest.
 The exchange occur via maximization of utility by each actor subject to a
constraint expressed in terms the given. The interests are
expressed in terms of structure in parametric role and the state
of the system is described in terms of distribution. So, they take
a structural form as given.
\par {\bf The Heider's  (1946,1958) balance theory.} Heider considered very simple relationship
 among two or three person. According to Heider, balance is not a real correction
of forces among elements but only perception of them by a person.
If one actor thinks that another actor treats him well, any
negative act "falls out" from hole picture. The cause is the
mental forces that aspire to restore equilibrium. From the
"behavioral" point of view the most important consequence of the
Heider and his followers is a supposition that: (1) positive
attitude is transitive ("I like that person, whom my friend like")
; (2) negative attitude is non-transitive (the following principle
does not work: "I hate the person whom my enemy hates").
\par Newcomb (1953) extends Heider's ideas to the system interaction in which the
standpoints of all interactants are jointly considered.
\par Cartwright and Harary (1956) formalized the concept of balance theory within a
graph theoretical setting. A circuit consisting of two or more
edges is a positive if circuit has an even number of negative
relations (in this model all relation may be only positive or
negative). A graph of relations is considered to be balanced if
all circuits consisting of two or more edges are positive.
Cartwright and Harary proved the important {\bf Structure
Theorem}:
\par The graph of relation is balanced if and only if it can be partitioned into
two subsets (one of which may be empty), such that all positive
cycles are within the two subsets and all of negative paths are
between them.
\par Davis (1967) generalized the structure theorem, shoving that balance holds, the
structure breaks up into a finite number of subgroups with
positive relationships within subgroups and negative relationships
between them. Davis suggested the next model of clustering: a
graph is clustered if no circuit includes exactly one negative
edge. In the model of ordered cluster Davis and Leinhardt (1972)
is supposed the existence of  a hierarchy of subgroups in which
every level includes at least one subgroup (in this model all
relation may be only positive or neutral).
\par Next very important problem is what kind { \bf mathematical tools} were used for
mathematical modeling of social processes. Cartwright and Harary
(1956) and Davis (1967) formalized the concept of balance theory
by using  a graph theory. Simon (1957) have used the differential
equations method. Harrison White (1963) worked out the abstract
algebraic group method for classification kinship system. Hunter
(1978) used the differential equations method. Fararo and
Scvoretz,  (so called E-state model, 1986) have used the
stochastic concept. Kovchegov (1987, 1994) worked out the
thermodynamics formalism for classification of stable social
system. Recently, the 'structural balance theory' research has
received renewed attention starting with Doreian and Mrvar (1996).
\par We can say that only Hunter (1978) tried to fulfill the Homans plan. He use
four social-psychological mechanisms: influence (my feeling toward
you will be more positive to the extent that my friends say nice
thing about you or to the extent that my enemies that my enemies
say bad thing about you), compatibility (my feeling toward you
will be positively affected if you say nice
 thing about my friends and nasty thing about my enemies), carryover (if I
like you, my feelings will tend to become more positive since I
will give you "the benefit of the doubt" on all ambiguous
statement. If I dislike you, everything will be just the
opposite), and reciprocity (my feelings will become more positive
if you say nice thing about me and I will react negatively when
you say negative things about me). But, as is easy shown, the
measure of  the non-balanced  invariant under Hunters' procedure
sociomatrices is greater or equal then measure of  balanced
sociomatrices.

 \par John von Neumann invented the {\bf cellular automata} in the late 40's and
 early 50's (1945, 1949,1951) when he sought to investigate the question of life's
 origin by trying to design a self-reproducing automaton. The theory of cellular
 automata (Burks, 1966, Codd 1968) is an example of a deterministic system with
 local interactions. Every automaton resides in an integer point of two or more
 dimensional real space and has a state (normally 0 or 1).  The system has an initial
 distribution of states. The dynamics of the cellular system is defined by the
 system of rules. The state of an automaton in a given moment of time t depends on
 the state of the neighbors in the previous moment t-1 (time is discrete). The rules
 are defined for all possible combinations of states of neighbors. The cellular automata
 are a homogeneous system: all automata use the same rules. Every rule is a prescription
 that assigns a particular value of state for the given automaton for every particular
 distribution of states of neighbors.
\par The theory of interacting systems emerged as a branch from the
theory of probability towards the end of the 60's.  Historically,
the first problems that motivated people to pay attention to
interacting particle systems were problems of statistical physics
particularly the Ising model (Glauber, 1969, Dobruchin, 1971). The
problem was to construct thermodynamics descriptions of the
evolution of systems where the states have the classic Gibbs
measures and then find the phase transitions.
\par Very soon, other sciences found similar problems to those found in the theory
of local interacting systems.  We will only list the most popular
models, where the theory of local interacting system is used.
Clifford and Sudbury, 1973 and Holley and Liggett, 1975 then did
elaborations to the voter model (behavioral science).  The local
interaction was defined on the integer points of n dimensional
space.  Every voter (Holley and Liggett interpretation for a voter
model) can support one of two political parties, denoted by 0 or
1.  A voter located on the integer points of n dimensional space
with non-zero probability get the position of spatial neighbors.
Clifford and Sudbury gave the following biological interpretation:
two populations denoted by 0 or 1, fight for territory. The voter
model state of the system then contains the field 0 or 1. \par The
{\bf thermodynamics description} is the description of systems
that uses the unit of {\bf measure as the state of system.}  Thus,
the thermodynamic state for the system is measure.  \par For the
voter model, there exist at least two invariant measures: the
delta measure on the system states in which all people are located
in the same position, and secondly, for 3 or more dimensional
space the existence of additional families of invariant measures.
The contamination model (Harris, 1974) is a model of the spread of
infections, where healthy people surrounded by contaminated people
get infections with non-zero probability.  A sick person recovers
with a non-zero probability as well. \par The profound
presentation of these models and other models are contained in
book Liggett (1985).
\par Kovchegov 1984, 1994 elaborated the
model of a human society using nets of automata with relation and
reactions. This model belongs to the class of local interacting
systems. The dynamics on the nets of automata with relation and
reactions was defined by interacting with neighbors, where
neighbors are neighbors on a graph of relations (not spatial). The
difference between the previous model and the current model of a
human society is the existence of relations and psychological
reactions that adjust the perception of a neighbors' state.  This
adjustment depends on the state of relations.  This difference is
crucial in that essentially the system is non-ergodic. The system
is called ergodic if all initial measures of the system converge
to a unique invariant measure.  The social system will be balanced
(in the social meaning) if the system has maximum invariant
measures (maximum taken for all possible relations or fields of
psychological reactions on the edges of the graph of relations).
The balanced social system satisfies the principle of maximal
non-ergodicity.  All fields of psychological reactions that
provide a balanced (in the social sense) social system are product
potential fields.  A product-potential system is a mathematical
analogue of the social balanced system. \par In the temporary
theory of automata and computer science, the term "automaton" is
used in formal systems that have internal sets and transform input
words into output words.  This automaton has input and output
alphabets, the set of internal states, and the set of rules
(programs).  The rules create current symbols for output words as
a function of the current input word and changes the internal
state. For deterministic automaton the choosing of a particular
rule depends only on the internal current state. For probabilistic
automaton this choice depends on the state and is performed
randomly.  \par We, however, do not use this term in the sense of
theory of automata. Our meaning of the term "automaton" is more
similar to the meaning of "cellular automaton" in the cellular
automata theory. There the "automaton" can perform a few actions
as well.  In this article the automaton can only perform two
actions: random choice and psychological adjustment. Our automaton
can randomly choose a neighbor, but cannot transform an input word
into an output word. \par So when we use the term "probability
automata" this means that we have only used the property to make a
random choice.  But we enable our automaton with some number of
additional actions for future modeling of collective actions. Part
four contains an example of this type of application.  The ability
to make a large number of different actions (not limited to
psychological adjustments) is the main reason why we use the word
"automaton" instead of the word "element" the way it is used in
the theory of local interacting systems.
\par  The automaton models that were done in part on the structural balance
theory from Heider, Cartwright and Harary, Davis and the cognitive
theory of Festinger. Festinger's cognitive dissonance theory
(Festinger,1957) allows us to formalize the process of motivation.
Within the framework of this model an attempt to formalize the
notion of dissonance - the main notation of Festinger's theory -
was undertaken.
\par  It is supposed that every automaton at every moment is in a certain
 state that belongs to the finite set of state. In addition, every
automaton is endowed with a set of  "psychological" reaction that
form an algebraic group of permutations of the finite of the
automaton state. According the dissonance theory each automaton
has the ability to make "psychological" adjustment it's perception
of a neighbor in the graph relation. The adjustment depend on the
feeling of the choosing automaton
 toward the chosen automata. The following principle was used: if my "enemy"
 is feeling well then I'm feeling bad, and if he/she feeling bad then
I'm feeling well. For the "friend" the relation is opposite.
\par  For this class of models a structural form is given. It means that a
structure (the graph of relation) was put in parametric role and
the state of the system was described in terms of distribution of
the states of actors. Then for every graph of relation (social
structure) was constructed the finite discrete Markov chain. Every
Markov chain has set
 of invariant measures. The following criterion for selection of the
balanced structures was chosen (the principal of maximum
nonergodicity )
\par  "We must select the structures where the associated  Markov
chains have the maximum number of stationary measures  (the
maximum taken over all structure of relation)".
\par It means that only group of actors with the structures of relation
satisfied to the principal of maximum nonergodicity will be
survival. We can find the groups of reaction when the selected
structures will be balanced according Cartwright and Harary or
Davis.
\par This article has three parts. Part one contains description of theory of
locally interrupted system with relations and reactions. The
potential networks were deduced from locally interrupting system
by using so call "principle of maximum nonergodicity".  The system
are balanced in social sense if the set of psychological reactions
on the graph of relation satisfy the principle of maximum
nonergodicity and this system of reactions are product potential
on the so call "two-steps" graph of relations. In real life
survive only relatively stable groups and we can observe only
stable groups that in social science call balanced. So reason why
social system (group) is stable based on hidden potentiality of
reactions: only social systems with potential system of reactions
(potential fields) are stable (balanced in social sense).
 \par The main problem what will prove in part two is problem of existence of potential
 fields. For this purpose we use method of smooth fields on the solid domain and
 product integrals. For of smooth fields will be write the system of infinitesimal
 equations (system of partial differential equations) that must be hold for all
 potential fields. The system of partial differential equations can be transformed
 into the system linear differential equation with one additional condition: the
 matrix-solution and field have to be anti-commutative pair. Then was found solutions
 of infinitesimal equations: the solution is any parameterizations of intersections
 of intersection of second-degree surface (set of matrices that $A^2$=E) and arbitrary
 plane. The set of solutions can be represented in few canonical forms. Then property
 to be potential will be checked by computer calculations.
  \par Then potential field on domain will be transformed into discrete potential marking on
embedded graph of relation by using product integrals. The finally
will be found system differential equations that
  transfer any initial fields of reactions into potential.
\par A locally interacting
process for the product potential system of relations can be given
by an algebraic representation of an process of multiplication on
the randomly chosen so call "control matrix". In part three we
found one to one maps between thermodynamic states of system (the
thermodynamic state for the system is measure) and so call "left
ideals" on the semi-group of control matrices. The ideal matrices
have a very important property: when an arbitrary
stochastic/control matrix is multiplied from the left by an ideal
matrix one obtains a left ideal matrix. So the set of left ideal
matrices is the termination set for our stochastic product process
(spatial Markov's chain). It means that once the system reaches
the termination set the process can never leave the termination
set. Thus the left ideal matrices play a crucial role in the
description of our process.
 \vskip .2in
\par { \bf Part 1. THE GENERAL THEORY OF THE LOCALLY INTERACTING NETS OF THE AUTOMATA. }
\vskip .1in
\par In the first part of paper we will describe the general model of locally interacting
network of automaton.
 \vskip .05in
\par \bf 1.1  Formulation of the automaton model of human groups .
\vskip .05in \rm
\par The next object is called the {\bf model of human groups with relations}

 U =$\{$ $\Gamma$=(A,B);Y and f:B $\rightarrow$ Y;$E_i$ and $G_i$,$\forall$ i $\in$ A;
$\psi_i$ :Y $\rightarrow$ $2^{G_i}$,$\forall$ i $\in$ A  ;  Z ,
$\Pi$ , Act : $\Pi$ $\rightarrow$ Z  $\}$
\par {\bf The graph of relations}
\par Automaton model of the human society is a network of automata connected
to each other by a graph of relations $\Gamma$ = (A,B)  ,where A
is a set of the vertices,B is a set of the edges encoding the
existence of relations among the members of the group.It is
supposed that $\Gamma$ is connected, directed finite graph without
loops and if (i,j) $\in$ B,then (j,i) $\in$ B.
\par {\bf Example. }
 A complete directed graph ${\Gamma}_3$ : A = $\{$ 1,2,3 $\}$ , B =$\{$(i,j),
 $\forall$ i,j $\in$  $\{$ 1,2,3 $\}$, i $\not=$ j $\}$.
\par {\bf A set of states of relations and a function of relations .}
Let Y be a set of states of relations among members of the group
and f = $\{$ f(i,j) ,for $\forall$ (i,j) $\in$ B,f(i,j) $\in$ Y
$\} $ is the structure of relations. The relations are symmetrical
if f(i,j) = f(j,i) $\forall$ (i,j)$\in$ B.
\par {\bf Example.}
 Y =$\{$ +1,-1 $\}$ is the set of states , $ y_{i,j} $ = f(i,j) = +1 if automaton
i has positive "relation" to automaton j and $ y_{i,j} $ = f(i,j)
= -1 if the "relation" is negative.
\par {\bf The set of states of automaton and a group of reactions.}
\rm It is supposed that every automaton at every moment is in a
certain state that belongs to the set of states E. In addition,
every automaton is endowed with a set of "psychological" reactions
that form an algebraic group G and are realized as transformation
E ( permutation for finite E ) on itself,
 i.e.  $\forall g \in G\>is\>g: E \rightarrow E $.
\par {\bf Example.}
 $ X_i$ = X =$\{ +1 , -1 \} $,where the state signed by symbol "+1" may be
interpreted as positive and "-1" - as negative.
\par Let $ G_i$ = G =$\{ g,e \} $ is group  and it is realized as substitutions
of the set of states X = $\{ +1,-1 \}$:
$$ g=\left(\begin{array}{cc}-1& 0 \\ 0 & -1\end{array}\right) \; and \; e =\left(\begin{array}{cc}+1& 0 \\ 0 &+1\end{array}\right)$$
i.e. $g(x) = -x $, $e(x) = x $, $\forall x \in X $ and $ g^2 =e$.
\par The last property ( $ g^2 = e$  ) is very important for  social
modeling . This property can be interpreted as a model of the law
of logic called the { \bf double negation }. It's the main reason
why we must demand that this property holds for all psychological
reactions. It means that for any algebraic group G this property
must hold, i.e. for all $ g \in G \; g^2 = e $ holds.
\par {\bf The choice function}.
\par Every automaton has the ability to make a "psychological" adjustment of
it's perception of a neighbor in the graph of relations. The
adjustment depends on the feeling of the chosen automaton towards
the other one. And it is given by the "choice" function
$\psi_i(f(i,j))\in G,for \;\forall i \in A \> and\;\> \forall
(i,j) \in B $.
\par {\bf Example.}
 $ \forall i\; \psi_i (-1) = g ,\; \psi_i (+1) = e $
\par Consequently if automaton i has bad feelings toward automaton j , i.e.
f(i,j) $<$ 0, which means he is "enemy" of j, then the state of
the second one is received by the first one after adjustment on
reactions as
  $ x_i =\psi_i (f(i,j))x_j =\psi_i (-1)x_j = gx_j = -x_j $. For the
 "friend" ( f(i,j)$>$ 0 ) the relation is opposite.
\par { \bf The set of actions and Act - function.}
\par Let Z be a set of actions ; the set Z has very complicate structure:
it contains, for example ,the actions ( operators ) which can be
done by one human being, by two, by three and so on. Every
operator has it's own domain . Let $ \Pi \; = \; \{ \; E_1, E_2,
\ldots, E_n \} $ be a partition of the set of states E ( it means
that next two equalities $ E \; = \; E_1 \cup E_2 \cup \ldots \cup
E_n \;  $ and $ E_i \cap E_j \; =  0  $ , hold for all $ i \;
\notin \; j $ ). The function $ Act : \Pi \rightarrow Z $ is
called the action function. If, for example, $ x \in E_i $, then
Act(x) is the set of possible actions ( operators ) from the set
of actions Z , which required i people for realization. In case
when our group is a discrete discontinuous  group and has the
fundamental domain we can transform an arbitrary  partition of the
fundamental domain to a partition of the whole E. The { \it normal
partition } is the partition which is generated by the partition
of the fundamental domain.
\par { \bf Example .} Let  G  = $\{$ f , g $\} $ be a group which is generated
by two elements f(z) = 2z and  g(z)  = $ { {3z + 4} \over { 2z +
3}} $ . The group G transforms the set  E  = $ H^2 $ = $ \{ $ z =
x + iy ,  where  x is  positive  real  number $ \} $ onto  $ H^2$.
We easily see that this group has a convex fundamental (the
hyperbolic quadrilateral) domain D and an arbitrary partition of D
generates the normal partition of E by the group of
transformations G.
\par Note that the M\"obius group G = $ \{$ f , g $\} $ is generated by four
reflections with respect to any side of the fundamental polygon
which contains exactly four sides ( all sides are circles ). But
for a reflection the double negation property holds, it means that
if we map our space $ H^2 $ two times by reflection ,then we
receive the identity function ( the identity function on $ H^2 $
is the function that maps each element of $ H^2 $ to itself ).
\par Suppose that the set of state of relationship Y is  a finite linearly
ordered  set and $ Y \; = \; \{ y_1 , y_2 , \ldots ,y_n \} $ ,
where $ y_i R y_j $ if and only if  i$<$ j . Let $ A(y_i) \; = \;
y_{i+1} $ , if $ i \leq n - 1 $ , $ A( y_n ) \; = \; y_n $ and $
B(y_i) \; = \; y_{i-1} $ , if i $>$ 1 , $ B(y_1) \; = \; y_1 $ .If
$ E \; = \; E_{-m} \cup E_{-m+1} \cup \ldots \cup E_{-1} \cup E_0
\cup E_1 \cup \ldots \cup E_{m-1} \cup E_m $ , then for $ x \in
E_k $ Act(x) = $ A^k $ and for $ x \in E_{-k} $  Act(x) = $ B^k $
,where $ 0 \leq k \leq m $ .It means that a person can make
relation with another person better or worse by operators A and B.
In this case our reactions $ g_{i,j} \in \Psi (y_{i,j}) $ , where
$ y_{i,j} \; = \; Act(x_i(t-1)) $ belong to the set of actions Z
too.

\par The \it state of a system \rm is the
function x = $\{ x_i$ , for all $i \in A $ and $ x_i \in E \} $.
Let W be the set of states of the system.
\par The marking of the edges of the graph $ \Gamma $ by reactions is given
by the set  R(G) = $\{$ $g_{ij}$, $\forall$ (i,j) $\i$ B $\} $. We
consider , as a rule , those marking R(G)  for which there exist a
function of relation f such that $ g_{ij}$ $\in$ $\psi_i$(f(i,j))
$\forall$ i $\in$ A  for any  j $\in$ $\partial \{i \} $ , where $
\partial \{ i \} $ is the set of neighbors of the automaton i on $
\Gamma $ . When it is easy to understand what kind of group is
used we will use the notation  R , i.e. the symbols R and R(G)
have the same meaning . At the same time we will use a "two-steps
graph" $ \Gamma^* $ : $ \Gamma^* \; = \; (A^* , B^*) $ , where $
A^* \; = \; A $ and $ (i,j) \in B^* $ if and only if there is $ k
\in B $ such that $ (i,k),(k,j) \in B $.
\par The marking of the "two-steps graph" $ \Gamma^* $ by reactions $ a_{i,l}
(k)$ = $g_{k,i}^{-1}g_{k,j} $ will be denote by symbol $ R^* $. It
means that $ R^* $ = $\{$ $a_{i,j}$(k)  , $\forall$ (i,j) $\i$
$B^*$ and for all admissible k , it means for all k that there is
a path  $ \{$ (i,k),(k,j) $\}$ $\subset$ $\Gamma$ $\} $.
\par Now we can define a "product integral" or P-integral along the way on the $ \Gamma
$ ( $ \Gamma^* $ ).
\par If $ L_{i,j} $ = $\{$ $(i_1,i_2)$,$(i_2,j_3)$ , $\ldots$ ,
($i_{n-1}$,$i_n$) $\} $ is an arbitrary directed way from i to j
on the graph $ \Gamma $ ( where $ i_1$ =i  , $i_n$=j  ), then a
product integral along the directed way $ L_{i,j} $ is $$
R(L_{i,j}) \; = \; g_{i_1,i_2} g_{i_2,i_3} \ldots g_{i_{n-1},i_n}.
$$
\par If $ L_{i,j}^* $ is an arbitrary directed way on the graph $ \Gamma^* $
from i to the j, $ L_{i,j}^*$ = $\{
(i_1,i_2,i_3)$,$(i_3,i_4,i_5)$, $\ldots$
$(i_{2k-1},i_{2k},i_{2k+1}) \} $, where (i,k,j) is the edge (i,j)
of the graph $ \Gamma^* $ that corresponds to the path $ \{$
(i,k),(k,j) $\} $ on the graph $ \Gamma $ and $ i_1$ = i , $
i_{2k+1} = j $ , then a product integral under the directed way on
the "two-steps graph" is $$ R(L_{i,j}^*) \; = \; a_{i_1,i_3}(i_2)
\ldots a_{i_{2k-1},i_{2k+1}}(i_{2k}). $$
\par { \bf Example } 1. Suppose we are given a complete, therefore not
bipartite , graph of relations $ \Gamma_3 $  and its  two-steps
graph $ \Gamma^* $ ( see figure 1 ).The way $ L \; = \; \{
(1,2),(2,3),(3,1) \} $ is the close directed way on the graph $
\Gamma_3 $ and way $ L^* \; = \; \{ (1,2,3),(3,1,2),(2,3,1) \} $
is the close way on the graph $ \Gamma^* $. The way $ L^* $ on the
graph $ \Gamma^* $ is induced by the way L on the graph $ \Gamma $
. We easily can figure out the product integrals R(L) and $ R(L^*)
$ : $$ R(L) \; = \; R( \{(1,2),(2,3),(3,1) \} ) \; = \;
g_{1,2}g_{2,3}g_{3,1} $$
$$ R(L^*) \; = \; R( \{(1,2,3),(3,1,2),(2,3,1) \}) \; = \;
a_{1,3}(2)a_{3,2}(1)a_{2,1}(3) $$
\par 2.Suppose we have the bipartite graph $ \Gamma $ and its two-steps graph
$ \Gamma^* $ . We easily see that a graph $ \Gamma^* $ has two
connecting components ( it's true for all bipartite graph  ) . Let
L = $ \{$ (1,2),(2,3),(3,4),(4,1) $\} $ be the close directed path
on the $\Gamma $ , and let $ L_1^*$ = $ \{$ (1,2,3),(3,2,1) $\} $
and $ L_2^*$ = $\{$ (2,3,4),(4,1,2) $\} $ be two different paths,
on the two different components of the graph $ \Gamma^* $ , that
were induced by way L.
\par Similarly we find the all multiplicative integrals : $$ R(L) \; = \;
g_{1,2}g_{2,3}g_{3,4}g_{4,1} \quad R(L_1^*) \; = \;
a_{1,3}(2)a_{3,1}(2) \quad R(L_2^*) \; = \; a_{2,4}(3)a_{4,2}(1)
$$
\par The dynamics of the states of the system is determined by family of
conditional probabilities $Q=\{q_{i}(x|x_j,\forall j \in
\partial\{i\}), \forall i\in A\}$.It is supposed that
$$q_{i}(x|x_j,\forall j\in \partial\{i\}) > 0 $$
if and only if $x\in [x_{j},\forall j \in \partial\{i\}] $ ,where
$ [x_1,\ldots,x_m] $  is non-ordered set, sublattice and so on if
E is a lattice.
\par If the family of conditional probability Q  is fixed, then on the set of
probability measures M a Markov operator
$$\mu Q(x) = \sum_{y\in F^{-1}(x)}\prod_{i\in A} q_{i}(y_{i}|g_{ij}x_{j},
\forall j\in \partial \{i\}) \eqno (1) $$ is defined for any
structure of relations f, where $\mu $ is an arbitrary probability
measure, $\mu \in M $. Here $F^{-1} =\{y \in W: x\in F(y)\} $,
$$F(x) = \prod_{i\in A} [g_{ij}(x_{j})\forall j \in \partial \{i\}] $$

\vskip 0.05in

\par{ \bf 1.2 The product potential ( non-dissonant ) system.}
\vskip .05in
\par Let $ Y = \{ y_{i,j} ,\forall (i,j)\in B \} $ be the structure of
relations. The system of reactions $ R =\{ g_{i,j} ,\forall (i,j)
\in B\;and\; g_{i,j} \in \psi_{i}(y_{i,j})\}$ is called {\it
potential ( non-dissonant ) } if there exists a state of the
system $ x \in W $ that $\forall i \in A $
$$ g_{i,j}x_{i} = g_{i,k}x_{k} \; ,\; \forall j,k \in \partial\{i\}$$ hold.
\par  Let $ W_0^{R} $ be the set of non-dissonant states of the system.
\par If $ F(W_0^{R})\subseteq W_0^{R} $ then the non-dissonant structure of
relations remains non-dissonant for some time.
\par   When  $ W_0^R $ is not empty and $ F(W_0^R)\subseteq W_0^R $ ?
\par { \bf The potential system in physics .}
\par The vector field $ \overline { F }(x,y) \; = \; ( \; P(x,y) \;, \;Q(x,y)
\; ) $ is called a potential if and only if $$ \int_{L_{A} }
P(x,y)dx \; + \; Q(x,y)dy \; = \; 0 \eqno(*) $$ holds for all
closed paths from a point A to point A.
\par If condition (*) holds, we can define a function ( { \it a potential
function } )
$$ u(B) \; = \; \int_{L^{1}_{A,B}} \; P dx \; + \; Qdy \; = \; \int_{L^{2}_{
A,B}} Pdx \; + \; Qdy $$ for any two paths from a fixed point A to
an arbitrary point B.
\par In this case the equalities $$ \overline{F}(x,y) \; = \; ({{ \partial u }
\over { \partial x }} , {{ \partial u } \over { \partial y }} ) $$
and du = P(x,y)dx + Q(x,y)dy hold.
\par In our case we have similar  situation.The "product integral "
$$ R(L_{i,i}) \; = \; e \; \; ( R(L^{*}_{i,i}) \; = \; e) $$  for any closed path  $L_{i,i} $
($ L^{*}_{i,i} $) on the graph $ \Gamma $ ($ \Gamma^{*} $ ) ( we
can compare this property ,which was called property A1 before
with property (*) ). The potential function is $$ u(k) \; = \;
r_{i,k} \; = \; R(L^1_{i,k}) \;= \; R(L^2_{i,k}) \quad (u(k) \; =
\; r_{i,k} \; = \; R(L^{1*}_{i,k}) \;= \; R(L^{2*}_{i,k}) )$$ ,
for the arbitrary paths $ L^1_{i,k} $  ($ L^{1*}_{i,k} $) and $
L^2_{i,k} $ ($ L^{2*}_{i,k} $) from a fixed vertex i to an
arbitrary vertex k on the graph $ \Gamma$ ($ \Gamma^{*} $) .
\vskip .05in
\par { \bf 1.3 The main system of the potential equations.}
\vskip .05in
\par The system of equations $$ R(L_{i,i}) = e \quad for \; all \; i \in A \quad
 (R(L_{i,i}^*) = e \quad for \; all \; i \in A ) $$  for any
closed path $ L_{i,i}$ ($ L_{i,i}^* $) in the graph $ \Gamma $ ($
\Gamma^* $) is called { \it the main system of the potential
equations }.
\par A solution of the main system is a list  $ \{ g_{i,j} \forall (i,j) \in
B \} $ that makes each equation a true statement when the elements
of the group G $ \{ g_{i,j},\forall (i,j) \in B \} $ are
substituted in the main system of equations.
\par { \bf Example of solutions for the main system of equations for $ \Gamma_3 $ }
\par Let $ a_{1,2} = g_{3,1}^{-1}g_{3,1} ,a_{2,3} = g_{1,2}^{-1}g_{1,3} ,
a_{3,1} = g_{2,3}^{-1}g_{2,1} $.
\par The main equation is $ a_{1,2}a_{2,3}a_{3,1} = e $.
\par This equation has the family of solutions:
\par $ g_{1,2} = x_1 ,g_{3,1} = x_2 , g_{3,2} = x_3 $
 $ g_{2,1} = x_3^{-1}x_2 x_1 x_3^{-1} x_2 $
\par $ g_{1,3} = x_1 x_3^{-1} x_2 x_1 x_3^{-1} $
 $ g_{2,3} = x_3^{-1} x_2 x_1 x_3^{-1} x_2 x_1 x_3^{-1} $,
\par where $ x_1 ,x_2$ and $x_3 $ are arbitrary elements of group G
( $ a_{1,2} =x_2^{-1} x_3 $ , $ a_{2,3} =x_3^{-1} x_2 x_1 x_3^{-1}
$, $ a_{3,1} =x_3 x_1^{-1} $ ).
\par This family of solutions is very interesting. They show
us that different automata play different roles in the society .
The automaton 3 can react arbitrary to the automata 1 and 2 ( $
g_{3,1} \; = \; x_2 $ , $ g_{3,2} \; = \; x_3 $ , where $ x_2 $
and $ x_3 $ are arbitrary independent variables ). The automaton 1
can feel himself free with respect to the automaton 2 ( $ g_{1,2}
\; = \;x_1 $ ) , but must adapt its behavior with regard to
automaton 3. The automaton 2 plays a central role in the process
of maintaining a potential ( non-dissonant ) property in the group
. The automaton 2 simulates a very adaptable flexible personality:
it has to adapt its behavior with respect to the other two
automata.
\par This interpretation means that a potential system of automata can't be
nonhomogeneous.
\par  So the within network of relations arises the problem of existence of potential networks.
This problem can get a complete solution for potential fields
(markings) with values in set of the 2 by 2 matrices. \vskip
0.05in
\par { \bf  The thermodynamics description of nets of automata with relations and group of
reactions: the principle of maximum nonergodicity for a finite
graph and a finite order group of reactions. }
\vskip .05in
\par It is natural to suppose that in reality those and only those group structures are observed
where the matrix of transition probabilities of the corresponding
Markov chain has a limit at $ t \rightarrow + \infty $.
\par {\bf Model 1.}
\par In model one ([33]) we have used an existence of limit for transition matrix of Markov's
chain as criterion for selection of balanced graph.
\par Let P = (p(x,y) for any two states of system x and y). Our criterion
was demanding of existence of limit $ \lim_{n \rightarrow \infty}
P^n$. The analysis of model [33] show as that there is connection
between existence of limit and stationary points of map F and
number of stationary measures (pP =p).
 \par The limit exist if and only if $W^R_0$ contains only stationary points of map F
 ($W^R_0$ = $\{$ x $\vert$ F(x) = x $\}$).
 \par The limit exist if and only if the marking of edges $
 \Gamma$ is product-potential field ($R(L_{i,i})$ = e for any close direct path $L_{i,i}$).
 \par The number of stationary measures of our Markov's chain
 reach maximum (for all possible relations) only for system of
 automaton for which exist limit of iterations of transition
 matrix.
 \par {\bf Example.} (1) Suppose we have a complete graph of relations $ \Gamma_3 $ .
\par Y = $ \{ +1,-1 \} $ is the of state of relations .Let f = $ \{y_{1,2} =
y_{2,3} =+1\;,y_{3,1} =-1\; , y_{i,j} = y_{j,i}\} $ be a structure
of relations, The group of reaction is G = $ \{ g ,e \} $ , where
g(x) = - x and e(x) = x , for all $x \in E $ = $\{$ +1 ,-1 $\}$ .
The choice function is $ \psi(-1)$  =
 g , $ \psi(+1)$ = e . Let R = $ \{g_{1,2} =g_{2,3} = e $ ,  $ g_{3,1} = g $ ,
$ g_{i,j} =g_{j,i}\} $ be the marking is accorded with the
structure of relations f by the choice function $ \psi $, i.e. for
all $ (i,j) \in B $  $ g_{i,j} \in \psi _i(f(i,j)) $ .
\par $ \forall  x=(x_1 ,x_2 ,x_3 )\in W $ has
$$ x\rightarrow F(x) =\{ex_2 ,gx_3 \}\times \{ex_1 ,ex_3 \}\times \{ex_2 ,gx_1\}=
\{x_2 ,-x_3 \}\times\{x_1 ,x_3 \}\times\{x_2 ,-x_1 \} $$
\par It is mean that for arbitrary state of system ($x_1$, $x_2$, $x_3$) $\rightarrow$
($x_2$, $x_1$, $x_2$) with probability $q_{1,2}$ $q_{2,1}$ $q_{3,2}$. Similarly ($x_1$,
$x_2$, $x_3$) $\rightarrow$ ($x_2$, $x_1$, -$x_1$) with probability $q_{1,2}$ $q_{2,1}$
$q_{3,1}$ and so on. So we have Markov chain with 8 by 8 transition matrix P = $\{$
$p_{x,y}$ for any states of system x and y $\}$. The set of states of system contains
8 elements.
\par Right now we will describe the very important set W of states that F maps
into one element set ($\vert$ F(x) $\vert$ = 1): W= $\{$ (x, -x, x) , (-x, x, -x) $\}$.
Really with probability 1 F:(x, -x, x) $\rightarrow$ (-x, x, -x) and F: (-x, x, -x)
$\rightarrow$ (x,- x, x). It means that transition matrix is oscillated matrix and
limit does not exist. For this system we have only one stationary measure.
\par (2) Let R = $\{$ $g_{1,2}$ = $g_{1,3}$ = g, $g_{2,3}$ = e, $g_{i,j}$ = $g_{j,i}$ $\}$.
\par So system is balanced and field is product potential.
\par In this case F: x $\rightarrow$ F(x) = $\{$- $x_2$, - $x_3$ $\}$ $\times$ $\{$ -$x_1$,
$x_3$ $\}$ $\times$ $\{$- $x_1$, $x_2$ $\}$. The solutions of
equation - $x_2$=- $x_3$ , -$x_1$ = $x_3$ ,- $x_1$ = $x_2$ is set
W. In this case W contains only one family of solution (x, -x, -x)
for any x from X. All this solutions are stable points: F(x,-x,-x)
= (x,-x,-x). So with probability 1 F: (x,-x,-x) $\rightarrow$
(x,-x,-x).
\par How easy to check in this case Limit exists and system has two stationary
measures (pP =p).
\par (3) Let R = $\{$ $g_{1,2}$ = $g_{1,3}$ = $g_{2,3}$ = g, $g_{i,j}$ = $g_{j,i}$ $\}$.
So system is not balanced and field is not product potential.
\par In this case F: x $\rightarrow$ F(x) = $\{$- $x_2$, - $x_3$ $\}$ $\times$ $\{$ -$x_1$,
- $x_3$ $\}$ $\times$ $\{$- $x_1$, - $x_2$ $\}$. The set W is set
of solution of system of equations $x_2$=- $x_3$ , -$x_1$ = -$x_3$
,- $x_1$ = -$x_2$ contains one family of solutions (x, x, x) for
all x from X. F(x, x, x) = (-x,-x, -x) for all x from X. So F(-1,
-1, -1)) = (-1, -1, -1) and F(-1, -1, -1) = (+1, +1, +1) with
probability one. It means that transition matrix is oscillated and
Limit does not exist.
\par In this case we have exactly one stationary measure.
\par Note. The behavior of map F on the set W determines the property of entire system.

\par {\bf Model 2.}
\par In model [35] for selection of balanced groups was used  "principle of maximum
nonergodicity". For this purpose we use "two-steps" graph and
demand potentiality only for field of reactions on the "two-steps"
graph" $\Gamma^*$ (see conditions A1 below). It is mean that that
initial system of reactions we do not demand to be potential.
Instead of demanding potentiality of initial field R we demand for
field R that condition A2 (see below) hold.
\par The marking of the edges of the graph $ \Gamma $ by reactions is given by
the set $R=\{g_{ij}$, for all $(i,j)\in B$ and where $ g_{i,j}\in
\psi_{i}(f(i,j))\} $. We say that the marking R satisfies
condition \par \bf A1 \rm if for any vertex i and for any closed
path $L_{i,i}^* $ in the graph ${\Gamma}^* $ the equation
$R(L_{i,i}^*) = e $ holds, where e is the unit of the group G.
\par We say that the marking R satisfies condition \par \bf A2  \rm if for any
vertex i such that $\partial \{i\} $ includes at least two
elements the equalities $g_{ij}g_{ji} = g_{ik}g_{kj} \forall j,k
\in \partial\{i\} $ hold.
\par In this case we call reaction $ a_i=g_{i,j}g_{j,i} $ the { \it
characteristic reaction } of the element i . If $ a_i = a $ for
all $ i \in A $ then we call the reaction a { \it characteristic
reaction } of the group.
\par This trick give
us ability find reasonable theorems for calculation of number
stationary measure by founding solution of some equations.
\par  The analysis of Model 1 has shown that the criterion for the selection of the
structures of relations according to the existence of the limit of
the transition probabilities matrix of the associated Markov chain
is the same as the criterion for selection of the structures by
the maximum number of stationary measures fore associated Markov
chains (the maximum taken over all structures of relations). This
last principle of selection, called \bf the principle of maximum
nonergodicity \rm , will be used here.
\par{\bf Proposition }. If for marking R the conditions A1 and A2  hold, then
 $ W_0^R $ isn't empty and $ F(W_0^R)\subseteq W_0^R $ and
   $W_0^R = W_0 = \{ x\in W : \vert F(x)\vert = 1 \} $, where $\vert$F(x)$\vert$ is the number of
elements in the set F(x).
\par  If $\Gamma $ is a graph of relations that is not
bipartite then $W_0 = \{z(t),\forall t \in E \} $, where  $z(t)
=\{ x_j = R(L_{j,i}^{*})t,\forall j\in A$ where $ L_{j,i}^* $ is a
path in $ \Gamma \} $, i is an arbitrary fixed vertex and $x_i = t
$. If $\Gamma $ is a bipartite graph then $A = A_1\cup A_2,A_1\cap
A_2 =\emptyset,\Gamma_i^* =( A_i,B_i^*) $ are two connected
components of the unconnected graph $\Gamma^* $,i = 1,2, and $W_0=
\{z(t,r),\forall t,r\in E\} $, where $z(t,r)=\{x_s =
R(L_{s,i}^{(1)*})t , \forall s\in A_1; x_k =
R(L_{k,j}^{(2)*})r,\forall j\in A_2 \} $;i,j are arbitrary fixed
vertices that accordingly belong to $A_1,A_2 ;L_{s,i}^{(1)*},
L_{k,j}^{(2)*} $ are the directed sequences on $\Gamma_1^* $ and
$\Gamma_2^* $ respectively ; $ x_i $= t ,$ x_j = r $.
\par Suppose, that there is a positive integer number m such that equality
$ g^m \; = \; e $ holds for all $ g \in G $. In this case the set
$$ Orbit( \{g
 \},x)\; = \; \{ x , gx , \ldots , g^{m-1}x \} $$ is called a $ \{ g \} -
orbit $ of element $ x \in E $ , where $ \{ g \} $ is a subgroup
of the G , which is generated by g .Let $ \overline{ x }^0 \; = \;
(x_1 , x_2 , \ldots , x_n ) $ be initial state of the net of
automata containing n automata. Let denote by the symbol $ E(
\overline{x}^0 ) $  the set that consists of all different
elements $ x_1 , x_2 , \ldots , x_n $ , i.e. $ E( \overline{x}^0 )
\; = \; \{ x_1 , x_2 , \ldots ,x_n \} $.
\par { \bf Example.} Suppose , $ \overline{ x}^0 \; = \; ( 1,2,1,1,1 ) $ , then
$ E( \overline{x}^0 )$ = $\{$ 1,2,1,1,1 $\}$ = $\{$ 1,2 $\} $ .
\par Let $ gE( \overline{x}^0)$ = $\{ g^kx $ for all $x \in E( \overline{x}^0)
$ and for all integer numbers k $ \} $ . There is a partition of $
gE( \overline{x}^0 )$ : $ \tau_g$ = $\{ Orbit( \{g \},x_i) ,Orbit(
\{g \},x_l)$, $\ldots$ ,$ Orbit( \{g \},x_s) $ , where $ gE(
\overline{x}^0)$ = $Orbit( \{g \},x_i) \cup Orbit( \{ g \},x_l)$
$\cup \ldots \cup Orbit( \{g \},x_s) $ and for $ i \not= j $
equality $ Orbit( \{ g \},x_i) \cap Orbit( \{ g \},x_j)$ = $
\emptyset $ holds.
\par The number of elements (orbits) in the partition $ \tau_g $ is denoted
by the symbol  $ Norbit(E( \overline{x}^0),g) $.
\par Let D(W) ($ D(W_0 )$) be the graph of transitions for Markov chain that
is constructed on the states of the systems W ($ W_0 $) .

\par {\bf Theorem A.} For any marking R that satisfy conditions A1 - A2  ,
i.e. for any potential system, the following conditions are true:
\par (1) for any $ x \in W $ on the graph of transitions D(W) there exists a
directed sequence of finite length from x to $ W_0 $;
\par (2) $ W_0 $ is the only set of essential states, that is to say
once the system enters this set it will never leave it.

\par {\bf Theorem B.}  1. Suppose, the graph of relations $\Gamma $ is not
bipartite. Let i be an arbitrary fixed vertex. Then, for any
marking R which satisfies the conditions A1 - A2 ( for any
potential system ) and for any  z(x)$\in M_0 $ the following
equalities hold : $ F(z(x)) = z(b_{i}x), F^{2}(z(x)) = z(a_{i}x)
$, where $b_i$ is a solution of {\it the characteristic equation }
 $$  v^{2} = a_i  $$
\par  2. If the graph of relations $\Gamma $ is bipartite and i,j are arbitrary
fixed vertices that belong to $A_1 $ and $ A_2 $ then for any
potential marking R that satisfies A1 and A2  and for any
$z(x,y)\in M_0$ ,  F(z(x,y)) = $ z(b_{i,j}y,b_{j,i}x)$,  where  $
b_{i,j} , b_{j,i} $  are solutions  of { \it the characteristic
equations } $$ v*w = a_i \qquad w*v = a_j $$ ( $ v = b_{i,j} , w =
b_{j,i} $ ). Note, that $ F^{2}(z(x,y)) =
 z(a_{i}x,a_{j}y) $
\par { \bf Example. } Let the state of automaton E be a finite set and
therefore a group of reactions be the group permutation of the set
E . Every permutation g is a product of disjoint cycles and the
number of the cycles in this decomposition is equal to $
Norbit(E,g) $.
\par  The dimension of the simplex of the stationary probability measures of
the Markov chain minus one is equal to
 the number of the disjoint cycles  in the decomposition of
permutation  v if $\Gamma$ isn't bipartite  or of permutation $
\sigma :(x,y)\rightarrow (wy,vx)$  if $\Gamma$ is bipartite.
\par According to { \bf a principle of maximum nonergodicity } we have to find
the solutions of the characteristic equations which maximize the
number of the cycles in decomposition of itself or maximize the
number $ Norbit (E,v) $ (the maximum is taken over all solutions
of the equation ).
\par {\bf Conclusion.} The most important case for is case when initial field of reactions
R is product potential. In this case all characteristic reactions $ a_i$ equal e. If field
R is product potential then field R* (defined on the "two-steps" graph) is automatically
product potential. From the principle of maximal nonergodicity best solution characteristic
equation $v^2$ = e (v*w =e and w*v= e) is identical permutation e. In this case F(z(x))= z(x)
(F(z(x, y))= z(x,y)) and number of stationary measures equal number stable points of
transformation F (see Model 1). It is mean that for product potentiality of initial field
of reaction set W is set of stable points of function F. So we have deduced all property of
 Model 1 from Theorem B and "principle of maximal nonergodicity".
\par Next will find the general solution to the existence problem for heterogeneous
potential systems.
\vskip .1in
\par {\bf  Part 2. THE SOLUTION OF THE EXISTENCE PROBLEM FOR TWO
DIMENSIONAL HETEROGENEOUS PRODUCT POTENTIAL SOCIAL SYSTEM  }
\vskip .1in
 \par {\bf  2.1. Set up of problem and definitions. }
\par  The social system or network is oriented graph (graph of
 relations), where any node represents the person and any edge
  represents relations. If any edge of social system marked by
  elements of algebraic group G (group of psychological reactions
   on the type of relations, where $g^2$ = e) and if product all
   elements along arbitrary closed path in the order in which the
    path goes equals unit element, then social system has called
     the social potential system or network.
\par Very attractive way to solve existing problem is just convert
 problem for discrete object (social network) to the similar
 problem on a rigid medium. Why? Because for the solid domain we
 can use calculus and mains concepts of theoretical physics:
 non-Abelian fields, infinitesimal equations for fields and so on.
\par The main problem that was solved in this article was problem
of the existence of social potential marking (fields). For this
purpose was created special method by using the smooth potential
fields on a rigid medium. For smooth potential fields we wrote the
system of infinitesimal equations that must hold for all potential
fields. It is a system of partial differential equations that was
transformed into the system of linear equations with one
additional
 condition on the solution: the matrix-solution and field have to
  be an anti-commutative pair.
\par Then we found solutions of infinitesimal equation: the solution
is any parameterizations of intersection of second degree surface
(set of matrices that $A^2$ =E) and arbitrary plane. The set of
solution can be represented in the few canonical forms. Then the
property of potentiality was checked by computer calculation for
all types of potential fields.
\par Right now we define the product or  path ordered integral for
the solid domain.
\par A path ordered integral for non-Abelian fields (P-integral) can be
defined as $$ P\left[ \int A dx\right] =  lim_{n -> \infty}
\prod_{i=1}^n(A(x_i) \delta x_i ) $$, where the product goes along
the path in the order in which the path goes.
\par The properties of P-integral see [31].
\par The physicist use the Pexp path ordered integral for non-Abelian fields:
  $$ Pexp{ \int A dx} = lim_{n -> \infty} \prod_{i=1}^n(1 + i A (x_i) \delta x_i )
  $$
, where the product goes along the path in the order in which the
path goes (see for instance [32]).
 \par  It is easy to see in our case that the definition of the integral
 depends from what the kind of integer n will be taken. If all number are
  even we get one result; for odd number we get completely different result:
  the determinant of A is -1 and the determinant of product the even (odd)
  number of matrices is +1 (-1). If we want to use P-integral as tool for
  theory of potential system, the number of step (n) must always be even
  integer number and we call it $P_2$ - integral. If number steps n is odd then we
  will receive $P_1$ - integral. We will use both.
\par {\bf  2.2 The example of solution of the existence problem for discrete potential system. }
\par We give the whole solution of this problem for group of reactions
(transformations) G =$\{$g, e$\}$, where $g^2$ = e and full graph
of relations. The solution will be done by algorithm. This
algorithm generate the family of G-potential fields (marks) or
potential fields $\{$ g(j,k), where j,k =1, $\ldots $ ,N $\}$ on
the full graph of relations with N notes, where N bigger than 3.
\par 1st step. Let us put g(1,2) equals g or e, g(1,3) equals g or e and g(2,3) = g(1,2)g(1,3).
               Let g(2,1) = g(1,2), g(3,1) = g(1,3), g(3,2) = g(2,3).
\par 2nd step. Let us put g(1,4) equals g or e and g(2,4) =g(4,2)= g(1,2)g(1,4).
                g(3,4) = g4,3) = g(1,3)g(1,4).
\par Step m. Let us put g(1,m) equals g or e, g(2,m) = g(1,2)g(1,m), g(3,m) = g(2,3)g(2,m)
 , $\ldots $ , g(j,m) = g(j-2,j -1)g(j-2,m), $\ldots $ , g(m-1,m) =g(m-2,m-1)g(m-2,m).
\par It is easy to prove that all this fields (marks) are potential and all potential fields
(marks) can be generated by this algorithm.
\par Examples. The mark g(1,k) = g(k,1) = g, where k =2, $\ldots $ , N and g(s,j) = g(j,s) = e,
 where s, j = 2, $\ldots $ , N. More interesting example: g(1,2)=g, g(1,3)=e,
g(1,4)=g,  and so on. Let g(2,3)=g, g(2,4)= e, g(2,5) =g and so
on. In general case g(k,k+1) =g, g(k,k+2)=e, g(k,k+3)=g,
g(k,k+4)=e, and so on.
\par {\bf  2.3. The general description of the field (marking)  A(x) and properties
of product N matrices. }
 \par For a heterogeneous potential system we have to find general description of
 the field (marking)  A(x), where $A(x)^{2}$ =e, A(x) belongs to group of
 psychological reactions for all x, and x is vector of parameters and when our group
 is the two dimensional group of the matrices. It is easy to see that the arbitrary
  2 by 2 matrix A satisfy this condition if and only if A= GZ$G^{-1}$, where G
  is arbitrary element of GL(2) and $$Z = \left(\begin{array}{cc}  0 & 1 \\1 & 0\end{array}\right). $$
\par It is also easily seen that $Z^{2}$=E and $A^2 $= GZ$G^{-1}$ GZ$G^{-1}$ =
 GZZ$G^{-1}$ = GZ$G^{-1}$ = A. For elements a(i,j) of the matrices A= GZ$G^{-1}$
 the next property hold: a(1,1) = - a(2,1) and a(1,2)a(2,1) = 1 -
$a(1,1)^{2}$. We then to deduce by direct calculation: a(1,1)= (bd
-ac)/(ad -bc), a(1,2) = ($a^{2}$ - $b^{2}$)/(ad -bc), a(2,1) =
($d^{2}$ - $c^{2}$)/(ad -bc), a(2,2) =(ac -bd)/ (ad -bc), where
      $$              G =\left(\begin{array}{cc} a &  b \\  c & d \end{array}\right), $$
where $\det$ G = ad - bc not equal zero.
 \par   This means that the set of 2-dimensional matrices that its square is
 unit matrix is a two dimensional manifold in three dimensional space D=$\{$ A(a,b,c)= $\{$
$\left(\begin{array}{cc}  a  & b \\ c &  -a\end{array}\right)$ , where  bc = 1- $ a^{2} $ $\}$.
\par Suppose we have matrix A(a,b,c), where bc = 1- $a^{2}$. How do we find matrix
G that A(a,b,c) =
 $G*Z*G^{-1}$?  For solving this problem we have to find solution the linear system
 A(a,b,c)G = GZ.
\par    Let  $ G =\left(\begin{array}{cc} x &  y \\ z &  w\end{array}\right)$, then solution $
G =\left(\begin{array}{cc}x  & ax + bz \\ z  &  cx  -az\end{array}\right) $ or $$ G =
\left(\begin{array}{cc} ay + bw &   y \\ cy - a w   &  w\end{array}\right). $$
\par So we can rewrite G = [x, Ax] or [Ay,y], where vector x (y) represent first
(second) column of matrix G.
\par The general solution is G = [x, Ax]  (G = [Ay,y]) for arbitrary vector x (y)
and inequality $\det$(G) not zero must hold. How easy to see
$\det$[x, Ax] =0 if and only if  x =0 or Ax = cx for nonzero x,
where c is constant. The eigenvalues of matrix A are +1 and -1 and
eigenvectors are
       $ e(1) = \left(\begin{array}{c}- b  \\ a -1\end{array}\right) $     , $    e(2) = \left(\begin{array}{c} -b     \\  a + 1\end{array}\right).$
\par So, the $\det$(G) is nonzero if and only if vector x (y) does not equal 0
(zero vector) or not to be proportional to eigenvector e(1) or
eigenvector e(2).
 \par   Then we have to find the commutator of matrix Z: Com(Z) = $\{$S: SZ = ZS $\}$.
 \par In two dimensional case the general element of Com(Z) can be represented in form
 S = a E +b Z, where a and b are arbitrary numbers and E is unit matrix. It means
 that $GZG^{-1}$ and $(GS)Z(GS)^{-1}$ are same for all S from Com(Z).
  \par   The product of two matrices A(1)A(2) can not belong to set
   $$      D = \{ A= GZG^{-1} \; for \; all \; G \; from \;
                                    GL(2)\} $$
 because $\det$ (A) = -1, but $\det$(A(1)A(2) ) =+1.
 \par Then A(1)A(2) can be equal unit matrix E if only if  A(1) = A(2). So, the product
 only odd number of elements from D can belong to D.
\par {\bf  2.4 The infinitesimal equations for social potential fields and solution
of the existence problem for general two-dimensional social
potential system on the solid set. }
 \par Let us define set of 2 by 2 matrices A(a,b,c) that square equal E as
 $A(a,b,s) = \left(\begin{array}{cc}a  &  b \\ c & -a \end{array}\right)$, where bc= 1- $a^{2}$.
\par Then take the square [0,1]x[0,1] on the plane xOy. Let h=1/n, where n is integer number,and
divide the square on the $n^{2}$ small squares with length of side
equal h. Then divide any small square on the two triangles by main
diagonal. Suppose we three smooth functions $f_1$(x,y),$f_2$(x,y),
 and $f_3$(x,y) defined on the square [0,1]x[0,1], where  $f_2$(x,y)$f_3$(x,y) =
 1 - $ f_1(x,y)^2 $. In this case we automatically define set of matrices A($f_1$(x,y),
 $f_2$(x,y), $f_3$(x,y))  on the square [0,1]x[0,1]. Let us take the triangle grid
 with step h and define the network of mark of edges by matrices A($f_1$(x,y),
 $f_2$(x,y), $f_3$(x,y)) taken in middle points of edge. We can start from interval
  y=0 and 0 $\le$ x $\le$ 1. We will find condition when heterogeneous (non-
homogeneous) distribution is potential. It means that product
integrals along two curves started in same point and ended in same
point are equal. It means that we have to find condition on the
$f_1$(x,y), $f_2$(x,y), $f_3$(x,y): M($c_1$(C,D), A($f_1$(x,y),
$f_2$(x,y), $f_3$(x,y))) = M($c_2$(C,D), A($f_1$(x,y), $f_2$(x,y),
$f_3$(x,y))) , where C and D are arbitrary points on the square
[0,1]x[0,1] and c1 and c2 are curves connected C and D.
\par {\bf Lemma.} The social fields with values in the set of matrices A($f_1$(x,y),
$f_2$(x,y), $f_3$(x,y)) is potential if and only if satisfy the
system of differential equations) :
  $$  A(d^2A/dxdy) +dA/dydA/dx =0 $$
\par , where $A^2$ = A and $f_2$(x,y)$f_3$(x,y) = 1 - $f_1(x,y)^2 $.
  \par {\bf Proof.} Let use next notation dA/dx =$ A_{x}$, dA/dy =Ay, $d^{2}A/dxdx$ = $A_{xx}$,
 $d^{2}A/dxdy $= $A_{xy}$, $d^{2}A/dydy$ = $A_{yy}$ and so on.
 \par We have to calculate the production of four matrices
      A(x, y +dx)A(x +dx, y+dy)A(x+dx,y)A(x,y) and equate the E. But A(x, y +dx)A(x +dx,
y+dy)A(x+dx,y)A(x,y) = E + (A$A_{x} $ + $  A_{x}$ A)dx + (A
$A_{y}$ + $A_{y}$ A)dy + ((1/2)A $A_{xx}$
 +A $A_{x}A_{x}$ A +  ((1/2)$A_{xx}$ A)$ d^{2}x$ + (A $A_{xy}$ + A $A_{y}A_{x}$A + $A_{y}$ Ax +
 $A_{y}$ A $A_{x}$A)dxdy + ((1/2)A$A_{yy}$ + $A_{y} A_{y}$ +  ((1/2)$A_{yy}$A) $d^{2}Y$ + $\ldots$ =
 E  + (A$A_{xy}$ + $A_{y}A_{x}$)dxdy + $\ldots$ . So for all dx and dy the equality
    E  + (A$A_{xy}$ + $A_{y}A_{x}$)dxdy =E hold. But it is possible if and only if A$A_{xy}$ +
 $A_{y}A_{x}$  =0.
 For calculation we have widely used next property: AA =E. It means that A$A_{x}$ = -$A_{x}$A,
 $A_{xx}$ A + 2$A_{x}$ $A_{x}$ + $AA_{xx}$=0, $A_{xy}$A +$ A_{x}A_{y}$ +$A_{y}A_{x}$ + A $A_{xy}$ =0
 and so on.
\par  The social potential marking (field) have to be solution to nonlinear partial
differential equation  A$A_{xy}$ + $A_{y}A_{x}$  =0 or $A_{xy}$ A+
$A_{y}A_{x}$  =0.
 The equation is actually the system of linear ODE:
 $(AA_{x})_{y}$ = A $A_{xy}$+ $A_{y}A_{x}$  and $(AA_{y})_{x}$ = A$A_{xy}$ + $A_{x}A_{y}$.
 So our equations are
      $$ (A_{x}A)_{y}=0 \, and \, (AA_{x})_{y} =0. $$
 \par The last equations mean that
  $$  A_{x}A  = B(x) \, and \, AA_{y} = D(y), $$ where B(x)  (D(y)) is matrix-function
  only from one variable x (y) and equality  AB = -BA     ( AD = - DA ) hold.
 \par We can get more information about the solid potential system if transform it by using
 exponential representation for A: A = exp(B). For this purpose we will use the spectral
 representation for A. We calculate A = Z(1) - Z(2) and arbitrary function f(x)
 f(A) =f(1) Z(1) + f(2)Z(2), where Z(1) =(1/2)(A + E), Z(2) =(-0.5)(A -E), Z(1) + Z(2)
 = E, $Z(1)^{2}$ = Z(1), $Z(2)^2$ = Z(2), Z(1)Z(2)
= Z(2)Z(1) = 0. If  A = exp(B), where B is unknown matrix, then
B=ln(A) = ln(1) Z(1) + ln(-1) Z(2), where ln(1) =0, ln(-1) = $\pi$
+ $2\pi$ki,  $i^2$ = -1 and k is arbitrary integer. It means that
 $$                  B =\pi Z(2).    $$
\par We now easily to calculate exp(B)=E - Z(2)$\cos$(p) Z(2) + $\sin$(p) Z(2)i  = Z(1) - Z(2) =A.
\par The matrices A(a,b,c) can be rewrite in more familiar form. Let put b =x + iy =z,c = x - iy
 =$\overline{z}$. In this case we get
 A(a,z) = A(a,x,y) = $\left(\begin{array}{cc} a  &  z  \\    \overline {z}  &  -a \end{array}\right) $,
 \par where $ a^{2}$ + $x^{2}$ + $y^{2}$ =1 (x=(b + c)/2, y = (b -c)/i2 and det A=1).
\par  We can represent the field as matrix dependent from one complex potential function:
            $$ A(z)  = \left(\begin{array}{cc} \sqrt{ 1- z \overline{z}}  &     z  \\
                   \overline{z}  &     - \sqrt{1- z \overline {z}}\end{array}\right)
                   $$
, where z = z(u,v)= x(u,v) + iy(u,v) is an arbitrary potential
function on the square [0,1]x[0,1].
 So we have the number potential system equal the number of potential function on the square.
\par {\bf  2.5 General and particular solutions of 2-dimensional infinitesimal equation. }
\par The system of differential equations (infinitesimal condition for potentiality)
     A$A_{xy}$ +$A_{y}A_{x}$ = $(A_{x}A)_{y}$=0 and  $(AA_{y)})_{x}$ =0 are really is first order
 system of ODE.
\par  What follow is the general description of all non-constant solution.   We will start
 from A$A_y$ = C(y),where C(y) is matrix-function only from one variable y and equality
 AC = - CA hold. So first of all we have to describe all matrices C that AC = - CA. Really we
 will describe a set of A(a,b,c) where AC = - C: the (a, b, c) must satisfy
equation
                                         2a$C_{1}$ + c$C_{2}$ +  b$C_{3}$=0
for any matrix $$ C= \left(\begin{array}{cc}C_{1}  &  C_{2} \\
                                C_{3} &  - C_{1} \end{array}\right).$$
\par It means that all triplet number $C_{1}$, $C_{2}$, and $C_{3}$ we can find solution by
solve system of algebraic equations:
                                $a^{2}$ + bc =1 and    2a$C_{1}$ + c$C_{2}$ +  b$C_{3}$ =0.
\par It means that (a, b, c) have to satisfy quadric equation:
     $ c^{2}(C_{2}^{2}$ + 2bc($C_{2}C_{3}$ + 2$C_{1}^{2}$) + $b^{2}C_{3}^{2}$ =
     4$C_{1}^{2}$.
 \par We can then prove by direct calculation the next theorem.
\par {\bf Theorem.}  For any triplets ($C_{1}$(y), $C_{2}$(y), $C_{3}$(y)) there are
solutions of equation A$A_{y}$ =C(y) that can be found as
parameterization of the intersection of hyperbolic $a^{2}$ + bc
-1=0 and plane 2a$C_{1}$ + c$C_{2}$ +  b$C_{3}$=0. Similarly we
can describe the set of solutions of equation $A_x$A = B(x).
\par Surprisingly, that to get the normal two variable fields at is easier use 3 by 3 matrices.
\par  {\bf Examples of solutions}. For instance, if $C_{1}$ equal zero
($a^{2}$ + bc =1, c$C_{2}$ +  b$C_{3}$=0) and $C_{2}C_{3}$ $>$ 0
then our equation represent the hyperbola $a^{2}$  -
$b^{2}C_{3}/C_{2}$ =1. Let us put
  a(t) = $\cosh(t)$ , b(t) =$ \sqrt {C_2/C_3} \sinh(t)$ .
\par In this case  $$  A = \left(\begin{array}{cc} \cosh (t)  &   \sqrt{C_2/C_3} \sinh(t)  \\
                              - \sqrt{C_2/C_3}\sinh(t)  &    - \cosh(t)
                                                  \end{array}\right)$$, where t = t(x,y).
$ A_{t}$ = $\left(\begin{array}{cc} \sinh( t)     &    \sqrt{C_2/C_3} \cosh(t)
\\ - \sqrt{C_2/C_3}\cosh( t)   & -\sinh(t)\end{array}\right)$ and $A_y$
=$A_{t}t_{y}$ .
 Therefor $$ AA_{y} =t_{y} \left(\begin{array}{cc} 0 &  \sqrt{C_2/C_3}  \\ \sqrt{C_{3}/C_{2}} & 0 \end{array}\right).
 $$
\par So our equation $AA_{y}$ = C, where  C = $ \left(\begin{array}{cc} 0 &  C_{2} \\     C_{3}    &
   0 \end{array}\right)$ can be transformed  into system
   $ t_{y} \sqrt{C_{2}/C_{3}}$ = $ C_{2} $ ,
  $   t_{y}\sqrt{C_{3}/C_{2}}$ = $ C_{3} $ or
it means that really we have one equation $t_{y}$ =$
\sqrt{C_3C_2}$.
\par And after integration t =$ \int \sqrt{C_{3}C_{2}} dy$ + R(x),  where R(x) is an
arbitrary function from x or constant ( $C_{2}$ = $C_{2}(y)$,
$C_{3}$ = $C_{3}(y)$ are function of y or constants).
 In case when $C_{1}$ equal zero and $C_{2}C_{3}$ $<$ 0 we have an ellipse and solution
 $$ A = \left(\begin{array}{cc} \cos (t)   &     \sqrt{ - C_{2}/C_{3}} \sin( t)
\\  - \sqrt{ - C_{2}/C_{3}} \ sin( t)   &    -\cos(t) \end{array}\right)$$,  where
t =$\int \sqrt{C_{3}C_{2}}$ dy + R(x).
\par Suppose $C_{1}$ is nonzero function or constant. In this case we have
$$ \left(\begin{array}{cc}-\cosh(t)+ \sinh(t) & L^{-1}(- \sinh(t)+ \cosh(t) )
\\  2L\sinh(t)  &     \cosh(t)- \sinh(t)\end{array}\right)$$ or $$ A =\left(\begin{array}{cc}
         \cosh(t) & L^{-1}\sinh(t) \\
         - L\sinh(t) &   -\cosh(t)\end{array}\right). $$
\par {\bf Numerical calculation}. We can use numerical calculation for checking of the potentiality of fields
 (see previous theorem and calculations). Suppose $$A(t) =  \left(\begin{array}{cc} \cos(t) & \sin(t)  \\
               \sin(t) &  - \cos(t) \end{array}\right),$$
 where  0 $\le$ t  $\le$ 1 and A(t) is the solution of an infinitesimal equation.
 The parameterization of interval [0, 1] can be define as function t =f(x), where x belong
 some interval [a, b]. The P-integral from potential field must be independent from
 parameterization:
 $P\left[ \int A dx\right]$ = $ P\left[ \int A dy\right]$
 for two arbitrary parameterizations t = f(x), a $\le$ x $\le$ b, f(a) =0, f(b) = 1,
 t =g(y), c $\le$ y $\le$ d, g(c) =0, g(d)= 1 or close P-integral must be equal E ( x goes
  from a to b and then y goes from d to c). We found by numerical calculation this property
  for different parameterizations: f(x) = $\sin$(x), 0 $\le$ x $\le$ $\pi$/2,  g( y)
= $y^m$, 0  $\le$ y $\le$ 1 , m =2, 3 and so on.
 \vskip 0.05in
\par { \bf 2.6  Final step: transformation continuous case into discrete.}
\par We can use D- fields and (E, D) - fields defined in solid space.
(E,D) - fields can be generated from D-fields by chose finite
number points and then "over blowing" points in domains with
values on the bounds of domain equal to value in given points.
Then we must just define field equal to the E for all internals
points of domains.
\par When product potential D-fields will be chosen we
have to immerse graph into solid space and, if we wont get (E, D)
- field,  "over blowing" same nodes of immersed graph. Then for
getting discrete values of field on the edges we have to take P1
(limit for sets of partitions of odd number points on edge) or P2
(limit for sets of partitions of even number points on edge)
product integral along all edges. But for every close path on the
graph the number edges where we were taken P1 integral must be
even! Because of E - fields are product potential for P2- product
integral. We can you different combination of P1 and P2 integral
along different edges. Some times this procedure gives us many
combinations of product potential system of marking of edges.
 \par { \bf Example.}
 \par For a triangle we can get only two combinations: for all three
 edges were take $P_2$ - integral and for two edges were take P1 - integral
 and for one age was take $P_2$ - integral.
  \par We define the function of relation f(k, j) = $\det$G(k, j),
  where G(k, j) equal P1 or P2 integral for D- or (E,D)- field along edge (k, j).
  It is clear that to us that number negative values of relation equal the number edges with
  negative determinant of reactions, but this number must be positive.
   \par { \bf So, all product potential systems automatically are balanced
   systems.}
 \par And for product potential system structural theorem for full graph of
 relation is true: for product potential system exist a maximum two antagonistic groups.

   \par How arbitrary D field can be transformed into product potential field?
   \par Mechanism look very easy for D - field defined on the solid domain.
   What does it mean to product potential (we use $P_2$ product integral)?
   It means that all values must belong to plane Aa + Bb + Cc =0 (where A=2$C_1$, B=$C_3$,
   C=$C_2$)in given domain.  So we have systems $a^2$ + bc= 1 and Aa + Bb + Cc
    =0, where (a, b, c) belongs to small domain, that really is ellipse or hyperbola. Then
   we make gradient system on the surface $a^2$ +bc=1 that has direction to curve
    - ellipse or hyperbola  ($a^2$ + bc= 1 and Aa + Bb + Cc =0).
 \par The our program has been realized. We can submerse the graph
 without intersection into N- dimensional space and then we can find the non-constant
 social potential fields defined in the same N-dimensional Euclidean area.
 \vskip .073in
\par {\bf Part 3. THE DYNAMIC OF PRODUCT POTENTIAL SOCIAL SYSTEM. }
\par    In part three we will define a set of special "control matrices" M and a
generating set of finite words $M^*$ in the alphabet M.
Simultaneously, we will define a set of stochastic matrices $M_s$
and a set of finite words $M^*_S$. Then we will define an product
process on the set of stochastic and control matrices. Our
stochastic process will be a left-side multiplier acting on a
randomly chosen control matrix. Then we will have to make the
final
 step for the description of the process: every finite product of matrices (words
 in alphabet M) has to be multiplied by a special matrix Rg in which all elements
  are psychological reactions. We will use a matrix representation for
  psychological reactions and Rg will be a matrix with matrix elements.
  Multiplication on the Rg will be defined in a special way too: we will use
  symbol * multiplication.
\par The set of words in an alphabet M  contains
some special element: the so-called  "left ideal" matrices.
Products of control matrices M generate all left ideal matrices.
The ideal matrices have a very important property: when an
arbitrary stochastic/control matrix is multiplied from the left by
an ideal matrix one obtains a "left ideal" matrix. So the set of
"left ideal" matrices is the termination set for our stochastic
product process. It means that once the system reaches the
termination set the process can never leave the termination set.
Thus the "left ideal" matrices play a crucial role in the
description of our process.
\par The knowledge of the structure of $M^*$ allows us to prove the main theorem for
a product potential system: there is a one to one connection
between distinct ideals and invariant measures.
 \vskip 0.15in
\par { \bf 3.1. The algebraic description of the dynamics of the states for potential systems. }
\vskip 0.05in
\par The dynamics of the states of a net of automata was determined previously [33-36] by a family of
conditional probabilities. But in the case when group of relations
is a connected graph we will find the algebraic description of the
dynamics of the states. In the models, which will be presented in
this paper, the dynamics of states are controlled by a subset of a
stochastic matrices so call set of "control matrices". The set of
stochastic matrices $M_s$  are matrices with the integer entries
equal to 0 or 1 and with sum of all rows equal one.
\par  Let M(B) is set of control matrices, where control matrices are stochastic matrices
with zero elements on the main diagonal ( Tr(A)=0) and all
elements less or equal to elements of adjacency matrix B of
original connected graph. Elements of M(B) we will call { \it the
control matrices }.
\par Let $M(B)^*$ is set of finite length words in alphabet M(B). So
$M(B)^*$ is set of words $A_m$ $A_{(m-1)}$ $\ldots$ $A_1$, where
$A_1$, $A_2$, $\ldots$ , $A_m$ is matrices from M(B). So "word" is
productions of m matrices for any not negative m. The sets of
words $M(B)^*$ is algebraic semigroups.
\par Matrix of reaction Rg presents the product potential system reactions on the complete
graph: g(k,l) is reactions on the edge (k,l) and g(s,s)=e, where e
is identical transformation (ex=x for all x). Potentiality on the
complete graph means that g(k,l)g(l,s) g(k,s) = e, g(k,l)g(l,s)=e
for any three nodes k, l, s. Therefore, g(k,l)g(l,s) g=g(k,s)
because of property of potentiality and gg=e.
\par We will show that local interacting process for product potential system of
reactions can be defined as algebraic presentation set of control
matrices into semigroup of transformations.
  \par The representation is map F of algebraic semigroup H into the group of liner
  matrices GL, where for any two arbitrary elements from H property F($h_1$$h_2$) = F($h_1$)F($h_2$)
  hold  (F($h_1$) and F($h_2$) are linear matrices).
 \par Right now we show connection between old
description of the dynamic and new one on the particular example.
\par Suppose all edges of graph of relations are marked by the elements belonging to the group of
   reactions G in accordance with the given system of reactions and the choice functions. For the
   arbitrary finite group of relations the system of elements marking a graph of relations can be
   written in the form of a square matrix. The dynamics of the system after n steps is determined
   by a product of n matrices applied to the initial state of the net, where every matrix is a
   *- product of the control matrix and the square matrix of the transformations (every
   entry in this matrix is a transformation of the state space). If we are given a system of the
   conditional probabilities on the * - product matrix, then the dynamics of the network's
   state will be defined.
\par Let A = ( $a_{i,j}$ , where $ 1 \leq i \; , \; j \; \leq n $ )
and B = ( $ b_{i,j}$ , where $1 \leq i \; , \; j \; \leq n \; ) .
$ The matrix A*B = ( $ a_{i,j}b_{i,j}$ , where $ 1 \; \leq \; i \;
, \; j \; \leq n $ )  is called a *-product of the matrices A and
B.
\par { \bf Example }. Suppose, the graph of relations $ \Gamma_n $ is
complete and n = 3. Let $ \overline{x}^0 \; = \; ( \; x_1 \; , \;
x_2 \; ,\; x_3 \; ) $ is the initial state of the  net and a
marking (potential field)  R  = $\{ \; g_{i,j}$ , where $(i,j) \;
\in B \; \} $ are given.
\par The few steps of the dynamics of the states of the system will be
described below.
\par { \it Step 1. } We have $ \overline{x}^0 \rightarrow F(\overline{x}^0) \;
= \; \{ \underline{g_{1,2}x_2},g_{1,3}x_3 \} \times \{ g_{2,1}x_1
, \underline { g_{2,3}x_3 } \} \times \{ \underline{ g_{3,1}x_1 }
, g_{3,2}x_2 \} $ . Suppose that the first automaton chose the
state of the automaton number 2 , second - automaton number 3,
third - automaton number 1 ( the options are underlined ) . It
means that $ \overline{x}^0 \rightarrow \overline{x}^1 \; = \;
(g_{1,2}x_2,g_{2,3}x_3,g_{3,1}x_1 ). $
\par Let  matrix $$ Rg \; = \; \left(\begin{array}{ccc} e & g_{1,2} & g_{1,3} \\
g_{2,1} & e & g_{2,3} \\ g_{3,1} & g_{3,2} & e \end{array}\right)$$ be the matrix
of transformations  and matrix $$ C_1 \; = \; \left(\begin{array}{ccc} 0 & 1 & 0
\\ 0 & 0 & 1 \\ 1 & 0 & 0 \end{array}\right)$$ be the control matrix which
reflects the options ( the first automaton chose the state of
automaton number 2 , second - automaton number 3 , third -
automaton number 1 ).
\par We easily see that $$  C_1*Rg \; = \; \left(\begin{array}{ccc} 0 & 1 & 0 \\
0 & 0 & 1 \\ 1 & 0 & 0 \end{array}\right) * \left(\begin{array}{ccc}e & g_{1,2} & g_{1,3} \\
g_{2,1} & e & g_{2,3} \\ g_{3,1} & g_{3,2} & e \end{array}\right)\; = \; \left(\begin{array}{ccc} 0 & g_{1,2} & 0 \\0 & 0 & g_{2,3} \\ g_{3,1} & 0 & 0 \end{array}\right) , $$
and $ C_1*Rg \overline{x}^{0t} \; = \; \overline{x}^{1t} $ ,where
$ \overline{x}^t $ denote the matrix transpose to the matrix $
\overline{x} $.We can check it directly
$$ C_1*Rg \overline{x}^{0t} \; = \; \left(\begin{array}{ccc} 0 & g_{1,2} & 0 \\ 0 & 0 &
g_{2,3} \\ g_{3,1} & 0 & 0 \end{array}\right) \left(\begin{array}{cc} x_1 \\ x_2 \\ x_3 \end{array}\right) \; =
\; \left(\begin{array}{cc} g_{1,2}x_2 \\ g_{2,3}x_3 \\ g_{3,1}x_1 \end{array}\right)\; = \;
\overline{x}^{1t} $$
\par We see that both the methods give us the same result.
\par { \it Step 2.} Second step is $ \overline{x}^1 \rightarrow \;
F(\overline{x}^1) \; = \; \{ g_{1,2}x_2^1 ,g_{1,3}x_3^1 \} \times
\{g_{2,1} x_1^1 , g_{2,3}x_3^1 \} \times \{ g_{3,1}x_1^1 ,
g_{3,2}x_2^1 \} $   and suppose the first automaton chose the
state of automaton number 3 , second - automaton number 1 , third
- automaton number 2 , that will be represented by the control
matrix $$ C_2 \; = \; \left(\begin{array}{ccc} 0 & 0 & 1 \\ 1 & 0 & 0 \\ 0 & 1
& 0 \end{array}\right). $$ ( Note that $ x_1^1 \; = \; g_{1,2}x_2 \; , \; x_2^1 \;
= \; g_{2,3}x_3 \; , x_3^1 \; = \; g_{3,1}x_1 $ ).
\par It means that on the second step the state of the system $ \overline{x}^1 $
was transformed into the new state of the system $ \overline{x}^2
\; = \;
 ( x_1^2 , x_2^2 ,x_3^2 ) $ , where $ x_1^2 \; = \; g_{1,3}x_3^1 \; , \;
x_2^2 \; = \; g_{2,1}x_1^1 \; , \; x_3^2 \; = \; g_{3,2}x_2^1 . $
\par Similarly to what we done in the step 1 we can find that $$
C_2*Rg \overline{x}^{1t} \; = \; \left(\begin{array}{ccc} 0 & 0 & g_{1,3} \\
g_{2,1} & 0 & 0 \\
 0 & g_{3,2} & 0 \end{array}\right) \left(\begin{array}{ccc} x_1^1 \\ x_2^1 \\ x_3^1 \end{array}\right) \; = \;
\overline{x}^{2t} \; $$ and $ \overline{x}^{2t} \; = \; (C_2*Rg)
\overline{x}^{1t} \; = \; (C_2*Rg)(C_1*Rg) \overline{x}^{0t} $ and
so on .
\par So we found that product potential system satisfy next very
important equality
$$ (C_2C_1)*Rg = (C_2*Rg)(C_1*Rg) $$
\par After n steps we will obtain the state of system $$ \overline{x}^{nt}
\; = \; (C_n*Rg)(C_{n-1}*Rg) \ldots (C_2*Rg)(C_1*Rg)
\overline{x}^{0t}
$$ , where $ C_1 \; , \; C_2 \; , \; , \ldots ,\; C_{n-1} \; , \; C_n $ are
randomly chosen control matrices.
\par We call semigroups A($M^*$)=($M^*$)*Rg=$\{$ w*Rg for all finite words from
$M^*$ $\}$ and A($M^*_s$)=($M^*_s$)*Rg=$\{$ w*Rg for all finite
words from $M^*_s $ $\}$  semigroups of
transformations(operators). The elements of matrices of
transformations are reactions. For instances it will be 2x2
matrices (see Part 2). \vskip .05in
\par { \bf  3.2. The representation of the subgroup and description of the states
dynamics.} \vskip .05in
\par The homomorphism $ \rho :M^* \rightarrow(M^*)*Rg $
is a representation of the semigroup $ M^* $ into the semigroup of
operators A($M^*$), where $ \rho(w)$=w*Rg, for any word w from )
$M^*$.
\par { \bf Example. }
 Let $ C_2 \; = \left(\begin{array}{ccc} 0 & 1 & 0 \\ 0 & 0 & 1 \\ 0 & 1 & 0  \end{array}\right)
\quad C_1 \; = \left(\begin{array}{ccc} 0 & 0 & 1 \\ 1 & 0 & 0 \\ 0 & 1 & 0  \end{array}\right)
\quad and \;C_2C_1 \;=\left(\begin{array}{ccc} 1 & 0 & 0 \\ 0 & 1 & 0 \\ 1 & 0
& 0  \end{array}\right)$
\par Therefore $ (C_2*Rg)(C_1*Rg)\;=\;\left(\begin{array}{ccc}0 &g_{1,2}& 0 \\
0 & 0 & g_{2,3} \\ 0 & g_{3,2} & 0\end{array}\right) \left(\begin{array}{ccc}0 & 0 & g_{1,3} \\
g_{2,1} & 0 & 0 \\ 0 & g_{3,2} & 0 \end{array}\right) \; $
 $ = \left(\begin{array}{ccc} g_{1,2}g_{2,1} & 0 & 0 \\
0 & g_{2,3}g_{3,2} & 0 \\ g_{3,2}g_{2,1} & 0 & 0 \end{array}\right)$
\par $ = \left(\begin{array}{ccc}e & 0 & 0 \\ 0 & e & 0 \\
g_{3,1} & 0 & 0 \end{array}\right)$.
\par We use potentiality of field $g_{i,j}$: $g_{1,2}g_{2,1}$  =e,
$g_{2,3}g_{3,2}$ =e, and $g_{3,2}g_{2,1}$ = $g_{3,1}$.
\par Contrariwise  $ (C_{2}C_{1})*Rg $ = $ \left(\begin{array}{ccc} 1 & 0 & 0 \\ 0 & 1 & 0 \\ 1 & 0 & 0 \end{array}\right)$
* $\left(\begin{array}{ccc} e & g_{1,2} & g_{1,3} \\
g_{2,1} & e & g_{2,3} \\ g_{3,1} & g_{3,2} & e \end{array}\right)$= $\left(\begin{array}{ccc} e &
0 & 0 \\ 0 & e & 0 \\ g_{3,1} & 0 & 0 \end{array}\right)$.
\par So $ \rho (C_2C_1)$ =$(C_{2}C_{1})*Rg$=($C_{2}$*Rg)
($C_{1}$*Rg) =$\rho(C_2)$ $\rho(C_1)$.
\par In general case very easy to prove that homomorphism $ \rho :M^* \rightarrow A(M^*) $
 ($ \rho :M^*_s \rightarrow A(M^*_s) $) is a representation
of the ring $ M^* $ ($ M^*_s$) into the ring of operators  (A($
M^* $) (A($M^*_s$) ).
\par We call set I { \bf left ideal } if for any word { \bf w} { \bf w}I
belongs to I  ({\bf w}I $\subset$ I).
\par The next theorem will be base for main Theorem 2.
\par {\bf Note.} All previous examples graph of relations are complete. But we can use this the
  algebraic method (the method of algebraic description of dynamic of the states for
  potential system) for arbitrary finite connected graph of relation marked by product
  potentials system of reactions. For this purpose we have to make our graph complete by
  adding new edges. Then we will mark new edges by reactions equal product of reactions
  along any path on the original graph started in source vertex of edges and ending in
  terminating vertex of edges. The set of control matrices contains only control matrices
  that less or equal to adjacency matrix of original graph. Similarly the entries of the
  reaction matrices Rg are reactions on original graph and new ones (built on the added edges).
  \par {\bf Example of extended matrix of reactions Rg for connected, but not complete graph of reactions. }
  \par Suppose we have graph of relation (A, V), where A =$\{$ 1,
  2, 3, 4 $\}$ and set of edges V = $\{$ (1,2), (2,1), (2,3), (3,2),(3,4), (4,3), (1,4), (4,1)
  $\}$. Suppose that filed of reactions is potential. It is mean
  that $ g_{i,j} g_{j,i}$ =e for all edges from V and $ g_{1,4} g_{4,3} g_{3,2} g_{2,1}$ =
  $ g_{1,2} g_{2,3} g_{3,4} g_{4,1}$ =e. We add two pairs edges $\{$ (4,2), (2,4), (1,3),
  (3,1)$\}$ and expand initial fields on the new edges keeping
  new field potential. It is mean that $ g_{2,4}$ = $ g_{2,1} g_{1,4}$ , $ g_{4,2}$ = $ g_{4,1}
  g_{1,2}$, $ g_{1,3}$ = $ g_{1,4}g_{4,3}$, $ g_{3,1}$ = $ g_{3,4}g_{4,1}$.
   \par The matrix of reaction is $$ Rg \; = \; \left(\begin{array}{cccc} e & g_{1,2} & g_{1,4}g_{4,3} & g_{1,4}\\
g_{2,1} & e & g_{2,3} & g_{2,1} g_{1,4} \\ g_{3,4}g_{4,1} &
g_{3,2} & e & g_{3,4}\\ g_{4,1} & g_{2,1} g_{1,4} & g_{4,3} & e\end{array}\right)
$$
  The number of invariant measures will be equal to number of ideals as well. So in general
   case we have to study the semigroup of words $M^*$(B) generated by set of control matrices
   M(B) , where all elements less or equal to elements of adjacency matrix B of
original graph.
\par { \bf Theorem 1 } 1. Suppose that the graph of relations is not bipartite.
The number of ideals of the semigroup $M^*(B)$ equals the number
of the vertices and every ideal is generated by one control matrix
$ I_k \;=\;\{ t_{k,j} =1\; \forall j \in \{ 1,2,\ldots ,n \}
\;t_{i,s} \;=\;0 \; \forall \;i \not=k $ and $ \forall s \} \; \;
k \in \{1,2,\ldots,n \}. $
\par 2.Suppose that the graph of relations is bipartite  and $ \Gamma \;=\;
(A,V)$ , where $ A\;=\; A_1\cup A_2,A_1 \cap A_2 \;=\;\emptyset
$.The number of ideals of the semigroup $ M^*(B)$ equals $ \left|
A_1 \right|\left|A_2\right | $ and every ideal is generated by two
elements  S(i,j) and $ S^{-}(i,j)$ , where $ i \in A_1 \; , \; j
\in A_2 $ and  S(i,j) = ( $ s_{k,l}$ ,where $ s_{i,l}$ = 1 , $i
\in A_1$ , for all $ l \in A_1 \; ; \; s_{m,j} $ = 1, $ j \in A_2
\; \forall m \in A_2 $ and ; all other elements are equal 0 ) ,
 $ S^{-}(i,j)$ = ( $ s_{i,l}$ = 1, $ i \in A_1 $ , for all $l \in A_2 \;
; \; s_{m,j}$ = 1 , where $j \; \in A_2$ , for all  $m \in A_1$
and all other elements are 0 ).
 \par {\bf The sketch of proof of the theorem 1.}
\par Suppose that the graph of relations $\Gamma$ = (A,V) is not
bipartite, where A = $\{$ 1, 2, $ \ldots $ , n $\}$. For arbitrary
vertex i we will find set of vertices $ A_1$ = A(i) that connected
with vertex i (for every j from $ A_2$ exist edge (j,i) from V) .
Then we find second set $ A_2$ = $\{$ j $\vert$ exist edge
(j,k),where k $\in$ $ A_1 $ $\}$ and so on. Throughout finite
number steps we will reach the set A. So we have chain $\{$ i $\}$
$ \leftarrow $ $ A_1$ $ \leftarrow $ $ A_2$ $\ldots$ $ \leftarrow
$ $ A_{m-1}$ $ \leftarrow $ $ A_m$ = $\{$ 1, 2, $\ldots$ , n $\}$.
Every control matrix maps A = $\{$ 1, 2, $\ldots$ , n $\}$ into
subset of A. Take matrix $C_1$ that transform $ A_m$=A into $
A_{m-1}$, then take control matrix that transfer  $A_{m-1}$ into
$A_{m-2}$ and so on and on the last step take control matrix $C_m$
that map $A_1$ into $\{$ i $\}$. The product $C_1$ $C_2$ $\ldots$
$C_{m-1}$ $C_m$will be equal $I_i$.
\par Similar procedure can be used for getting ideals $\{$ S(i,j), $ S^{-}(i,j)$
$\}$ when graph of relations is bipartite ($ A$=$A_1$ $\cup$ $
A_2$. Only one difference: we will start from arbitrary pair
(i,j), where i belongs to $A_1$ and j belongs to $A_2$.
\par { \bf Representation of the semigroup  and dynamics of  a
state of a potential system. Graph of transitions for the Markov
chain.}
\par Now we are ready to describe of the dynamics of a state of a potential
system.
\par For describing the dynamics of a state of potential system the rings
$M^*$ and A($M^*$) and  representations $ \rho : \; M^{*}
\rightarrow A(M^*) $ will be used.  The dynamics of the states can
be realized as representation of a random product process on the
semigroup $M^*$ into the semigroup A($M^*$). It means that first
of all we get randomly chosen initial ward (matrix) $C_0$ from
$M^*$. Then we multiply initial word (matrix) on the randomly
chosen matrix from M and so on. All steps are independent.
Throughout n steps we get random trajectory $C_0$ , $C_1C_0$ ,
$C_2C_1C_0$ , $\ldots$ , $C_nC_{n-1}$ $ \ldots$ $C_1C_0 $ as a
result of the random product process on the semigroup $M^*$ . Then
we map our trajectory on the semigroup $M^*$ into the ring
A($M^*$) and get real trajectory $$ \rho(C_0) , \rho(C_1C_0) ,
\ldots , \rho( C_nC_{n-1} \ldots C_1C_0 ) $$ , where $
\rho(C_nC_{n-1} \ldots C_1C_0 )$ = $ (C_n*Rg) (C_{n-1}*Rg) \ldots
(C_0*Rg)$ = $ \rho(C_n) \rho(C_{n-1}) \ldots \rho(C_0 )$.
\par For an arbitrary initial word (matrix) with probability one we have to reach
one of the ideals of the ring $M^*$. Therefore, according to
theorem 1 we must examine the behavior of our system in the
ideals. It hardly simplifies our problem.
 \par { \bf Theorem 2. } The number of ideals of the semigroup A($M^*$)
 equals to the number of ideals of the semigroup $M^*$.
 \par {\bf Proof of the theorem 2.}
 The theorem 1  means that stochastic product process on
 $M^*$  with probability one converges to one of n ideals for non bipartite graph or
 $ \left|A_1 \right|\left|A_2\right | $ ideals for bipartite graph. \par Suppose we have non bipartite graph
 of relation.  We map our trajectory that converge to ideal $I_k$ into ring A($M^*$). The map
$\rho$ is representation and image of ideals $I_1$, $\ldots$ ,
$I_n$ are ideals in $\rho$(A($M^*$)): $\rho$({\bf w})$\rho$(I(k))
= $\rho$({\bf w}I(k))= $\rho$(I(k)). It means that $\rho$(I(1),
$\ldots$ , $\rho$(I(N) are left ideals of $\rho$(A($M^*$)).
  The ideal $ \rho(I(k)) $ can be reached for finite number of steps. The ideals $ \rho(I(k)) $
 represent final classes and system has to reach one of N final classes and never leave them.
 \par The prove of theorem 2 for bipartite graph absolutely similar to previous one.
\vskip .01in
\par {\bf Connection between ideals of semigroup A($M^*$) and number of stationary measure
of Markovs chain defined on the system's states.}
\par Suppose we have product potential system and let denote by symbol Z a set of
system's states. It is to easy to see that the set of final states
W of Markov chain defined on the Z is union of set {\bf I}Z for
all ideals of semigroup A($M^*$) (W =$\bigcup$ {\bf I}Z for all
 ideals A($M^*$)).
\par The number of stationary measures equal the number of elements of final state W
(see Part 1). \vskip .01in
\par { \bf Example (See Part1, Model 1, example 2) } Suppose we have product potential
field f = $\{$ $g_{1,2}$ = $g_{1,3}$ = g, $g_{2,3}$ = e and
$g_{i,j}$ =$g_{i,j}$ $\}$ defined on the complete graph of
relation $\Gamma_3$. Suppose that X= $\{$ -1, +1$\}$ and e, g: X
$\rightarrow$ X, where gx = -x and ex=x for all x.
\par In this case semigroup A($M^*$) has three ideals $I_1$*Rg
=$\left(\begin{array}{ccc} e & 0 & 0 \\ g & 0 & 0 \\ g & 0 & 0 \end{array}\right)$, $I_2$*Rg
=$\left(\begin{array}{ccc} 0 & g & 0 \\ 0 & e & 0 \\ 0 & e & 0 \end{array}\right)$, $I_3$*Rg
=$\left(\begin{array}{ccc} 0 & 0 & g \\ 0 & 0 & e \\ 0 & 0 & e \end{array}\right)$, where $I_1$
=$\left(\begin{array}{ccc} 1 & 0 & 0 \\ 1 & 0 & 0 \\ 1 & 0 & 0 \end{array}\right)$, $I_2$
=$\left(\begin{array}{ccc} 0 & 1 & 0 \\ 0 & 1 & 0 \\ 0 & 1 & 0 \end{array}\right)$, $I_3$
=$\left(\begin{array}{ccc} 0 & 0 & 1 \\ 0 & 0 & 1 \\ 0 & 0 & 1 \end{array}\right)$ are ideals of
$M^*$, and $ Rg \; = \; \left(\begin{array}{ccc} e & g & g \\ g & e & e \\ g &
e & e\end{array}\right)$  is transformation matrix. The set of system's states Z
is 8 elements set $\{$ $\left(\begin{array}{c} x_1 \\ x_2 \\ x_3 \end{array}\right)$, for all
$x_1$, $x_2$, $x_3$ $\in$ X= $\{$ -1, +1$\}$ $\}$.
\par All three set ($I_1$*Rg) Z, ($I_2$*Rg) Z, ($I_3$*Rg) Z are
identical sets (($I_1$* Rg) Z $\equiv $ ($I_2$*
 Rg)Z $\equiv $ ($I_3$* Rg) Z).
For instance ($I_1$*Rg) Z = $\{$ $ \left(\begin{array}{ccc} e & 0 & 0 \\ g & 0 &
0 \\ g & 0 & 0 \end{array}\right) \left(\begin{array}{c} x_1 \\ x_2 \\ x_3 \end{array}\right)$ $\}$ $\equiv $
$\{$ $\left(\begin{array}{c} x_1 \\ -x_1 \\ -x_1 \end{array}\right)$  = $x_1 \left(\begin{array}{c} +1 \\
-1 \\ -1 \end{array}\right) \; for \; arbitrary \; x_1 from X $ $\}$ $\equiv $
$\{$ $\left(\begin{array}{c} +1  \\ -1 \\ -1 \end{array}\right)$, $\left(\begin{array}{c} -1 \\ +1 \\ +1
\end{array}\right)$ $\}$ is two element set and W = ($I_1$*Rg) Z $\bigcup$
($I_2$*Rg) Z $\bigcup$ ($I_3$*Rg) Z $\equiv $ ($I_1$*Rg) Z
contains exactly two elements.
\par So Markov chain defined on the system's states has exactly two stationary measures
(or two thermodynamic states).

\par {\bf  Conclusion. }
\par A very natural assumption of social systems being locally interacting systems
with relations and psychological reactions together with the
principle of maximal nonergodicity enable us to show that a
product potential system is a balanced (stable) system in the
social sense.
 \par This principle of maximal nonergodicity selects
exactly product potential systems out of all locally interacting
systems of automata with relations and reactions. Maximal
non-ergodicity means that only systems (networks of automata) with
this property have a maximum possible number of states (the state
in thermodynamics is defined by the measure) for all possible
relations. So the principle of maximal nonergodicity can be called
the principle of maximal freedom for a group.
\par The system are balanced in social sense if the
set of psychological reactions on the graph of relation satisfy
the principle  of maximum non-ergodicity and this system of
reactions are product  potential on the so call "two-steps" graph
of relations. In real life survive only relatively stable groups
and we can observe only stable groups that in social science call
balanced. So reason why social system (group) is stable based on
hidden potentiality of reactions: only social systems with
potential system of reactions (potential fields) are stable
(balanced in social sense). This conclusion is not a big surprise
for natural (physical) systems. For instance the system consisting
of a star and a single planet is stable because the gravitational
interaction is potential (friction is absent). The potentiality of
gravitational interaction means that work done along any closed
path is zero. But for social science and, particularly for human
groups, a similar property comes as a surprise.
\par Therefore, the reason why balanced groups (systems) can exist forever is a
hidden potentiality of human reactions inside balanced groups.  This is the main
reason this system was studied in this article.
\par The main difficulty was the problem of existence for heterogeneous product
potential systems. This problem has been solved by constructing infinitesimal
differential equations. The problem of existence was completely solved for a smooth
potential by solving infinitesimal differential equations. All product potential
fields on the solid domain were found as solutions of infinitesimal differential
equations for two-dimensional matrices.
\par Then a graph of relations was embedded in the domain.  We proceeded by
describing all product potential marks on the graph of relations
by integrating continuous product potential fields along the
graph's edges. This procedure used two different types of product
integrals: $P_1$ and $P_2$.
\par We proved that any product potential system on the graph is a balanced system
in the social sense.  Therefore, we can say that in our model
being balanced is identical to it being a product potential
system. So the structural theorem for a balanced group (stating
that a balanced group has at most two antagonistic subgroups) is
true for all product potential systems containing full graphs of
relations.
\par We have proved that a locally interacting process for a product potential
 system of relations can be represented as an process of multiplication
 on a randomly chosen control matrix with a transformation into the original process
 in the end.  The main property of a product potential system concerning the number
  of stable measures (there is a one to one connection between ideals and invariant
  measures) was derived.
   \par This article proposes a bridge between certain concepts of natural
 sciences and sociology.

\bibliographystyle{amsplain}

\end{document}